\documentclass{article}
\usepackage{spconf,amsmath,epsfig}
\usepackage{multirow} 
\usepackage{enumerate}
\usepackage{mathrsfs}

\usepackage{amsmath} 
\usepackage{amssymb}

\let\OLDthebibliography\thebibliography
\renewcommand\thebibliography[1]{
  \OLDthebibliography{#1}
  \setlength{\parskip}{0pt}
  \setlength{\itemsep}{0pt plus 0.3ex}
}

\begin{document}\sloppy

\def\x{{\mathbf x}}
\def\L{{\cal L}}

\title{Domain-Aware Cross-Attention for Cross-domain Recommendation}
%
\name{Yuhao Luo\textsuperscript{1,2}, Shiwei Ma\textsuperscript{2}, Mingjun Nie\textsuperscript{2}, Changping Peng\textsuperscript{2}, Zhangang Lin\textsuperscript{2}, Jingping Shao\textsuperscript{2}, Qianfang Xu\textsuperscript{1}}
\address{\textsuperscript{1} {luoyuhao, xuqianfang}@bupt.edu.cn\\
\textsuperscript{2} {luoyuhao9, mashiwei, niemingjun, pengchangping, linzhangang, shaojingping}@jd.com}

\maketitle

\begin{abstract}
Cross-domain recommendation (CDR) is an important method to improve recommender system performance, especially when observations in target domains are sparse. However, most existing cross-domain recommendations fail to fully utilize the target domain's special features and are hard to be generalized to new domains. 
The designed network is complex and is not suitable for rapid industrial deployment. Our method introduces a two-step domain-aware cross-attention, extracting transferable features of the source domain from different granularity, which allows the efficient expression of both domain and user interests. In addition, we simplify the training process, and our model can be easily deployed on new domains. We conduct experiments on both public datasets and industrial datasets, and the experimental results demonstrate the effectiveness of our method. We have also deployed the model in an online advertising system and observed significant improvements in both Click-Through-Rate (CTR) and effective cost per mille (ECPM).
\end{abstract}

\section{Introduction}
\label{sec:intro}
In major e-commerce platforms such as JD.com and Amazon, precise prediction of Click-Through Rates (CTR) is crucial to increasing business revenue. 
These platforms encompass multiple business domains, each representing a specific display location or activity page, and are dedicated to presenting  attractive items to users, whether through mobile applications or PC websites. 
With the enrichment of features and the complexity of model structures, deep CTR models have made significant progress in recent years. 
However, traditional recommender systems typically engage in tasks within a single domain, utilizing data collected from a specific domain to train models and serve for a specific task. 
With the increase of new domains and the influx of new users, some users have insufficient interactions in these domains, presenting recommender systems with the challenge of cold starts.

Fortunately, the missing interactions of these cold-start users can be captured in other online domains, offering a solution to the cold start dilemma. 
For instance, a user may exhibit sparse interactions in one business domain but possess a rich history of clicks in another. 
Cross-Domain Recommendation (CDR) methods have been developed to address the challenges of cold-start users by transferring knowledge from relatively data-rich source domains.
By leveraging user interactions from source domain, these methods compensate for the data scarcity in the target domain, thereby improving the accuracy and personalization across various e-commerce business domains. 



According to the general cross-domain recommendation settings, domains are usually divided into source domain and target domain. The source domain typically has richer interactions. 
This paper primarily investigates how to utilize data from the source domain to enhance the quality of recommendations for cold-start users in the target domain.


Most existing cross-domain recommendation methods employ embedding mapping methods, operating under the assumption that user preference relationship between the source and target domains are similar, which resulted in transferring user embeddings from the source to the target domain through shared preference transfer methods \cite{EMCDR, SSCDR, TMCDR}. Some approaches argue that the complex relationship between user preferences of source and target domains vary from user to user, which leads to personalized preference mappings \cite{PTUPCDR}. however, in reality, these methods often overlook the differences in data distribution, item features, and scenario characteristics between the source and target domains. They fail to take into account the actual characteristics of the target scenario, which may lead to negative knowledge transfer.

To address this issue, we introduce a two-step domain-aware cross-attention network. From a practical application standpoint, to align with the recommendation tasks in the target domain and effectively integrate source domain information, this network extracts key behavioral sequence features from the source domain during the cross-domain transfer process. 
In addition, this attention network considers both coarse-grained and fine-grained attention on domain-level and item-level attention to selectively control what to transfer from the source domain, thereby reducing the likelihood of negative transfer.
Furthermore, most mainstream cross-domain methods require pre-training embeddings in both source and target domains. This is followed by training a mapping function to project users from the source to the target domain and make recommendations. 
In contrast to these cross-domain approaches, our method trains a model end-to-end. This greatly facilitates industrial training and deployment. For a newly added domain, we can fine-tune a suitable model on top of the existing one, avoiding training a brand new one completely. This approach is more conducive to the industrial maintenance of multiple domains and the rapid deployment of new ones.

The main contributions of our work can be summarized as follows:
\begin{itemize}
    \item To Cold Start problem in CDR, we propose a novel method called Domain-Aware Cross-Attention for Cross-Domain Recommendation (DACDR). This method utilizes a two-step cross attention network to capture transferable knowledge of each user from source domain to alleviate negative transfer problem.
    \item Our model is industrial-friendly as we simplify the model architecture and training steps to alleviate the gradient imbalance issue associated with using behaviors as inputs for the Meta Network. Inaddition, for new domains, we can rapidly fine-tune the existing model and deploy it online. 
    \item We conduct extensive experiments on both industrial and public cross-domain datasets to demonstrate the effectiveness and robustness of DACDR. The DACDR method has been deployed on an online recommendation system and achieved a gain of on both CTR and ECPM.
\end{itemize}

\begin{figure*}[ht]
    \centering
    \includegraphics[width=1\linewidth]{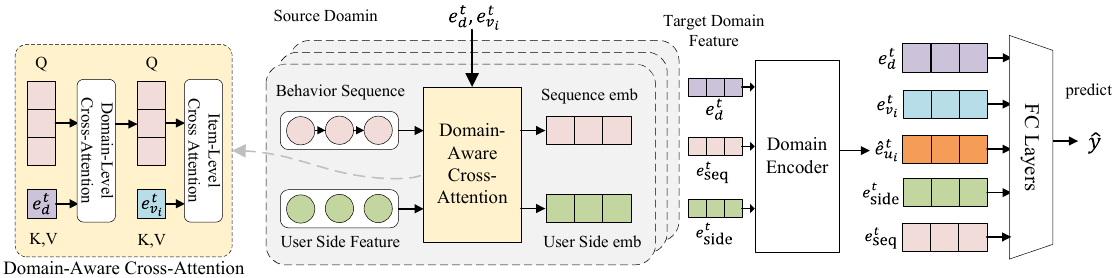}
    \caption{
    The overall architecture of DACDR. 
    }
    \label{fig:method}
\end{figure*}

\section{Related Work}
\subsection{Cross-domain Recommendation}
Cross-domain recommendation (CDR) aims to improve performance or minimize the number of labeled examples required in a target domain with the help of auxiliary domains. This methodology has garnered widespread attention in recommendation systems to alleviate data sparsity and the cold start problem in the target domain.
Initially, Collective Matrix Factorization (CMF) \cite{CMF} is a classic method that assumes all domains share a global user embedding matrix, simultaneously factorizing matrices from multiple domains.

In recent years, deep learning-based models have been proposed to enhance knowledge transfer. For instance, CoNet \cite{CoNet} utilizes cross connections between feedforward neural networks to transfer and integrate knowledge. DDTCDR \cite{ddtcdr} has developed a novel latent orthogonal mapping to extract user preferences across multiple domains while preserving user relationships across different latent spaces.

Another set of CDR methods focuses on constructing bridges for user preferences across different domains, which is most pertinent to our research. For example, EMCDR \cite{EMCDR} uses matrix factorization to generate latent factors for users and transfers user latent vectors via a linear function as the mapping function, while PTUPCDR \cite{PTUPCDR} learns personalized bridges for each cross-domain user. Our research falls into this bridge-based category but is distinct in approach. However, the aforementioned methods do not sufficiently consider the contextual features and data distributions of the target domain, and the approach of transferring preferences based on shared users may not fulfill the business requirements of the target domain in practical recommendation systems.

\subsection{Multi task / multi domain Recommendation}
In practical business domains, data for recommendation scenarios are often dispersed and unevenly distributed, posing challenges for model training and maintenance. Training individual models for each scenario not only adds to the complexity of maintenance but also yields unsatisfactory performance in data-sparse situations.
Some pioneering endeavors have attempted to employ multi-domain and multi-task learning as the most representative methodologies. Multi-task and multi-domain recommendation segment data into groups based on tasks or domains and utilize well-designed structures to learn parameters that are specific to each group.

In the realm of multi-task learning, the Share-Bottom structure \cite{ShareBottom} is proposed to share the input embedding layer at the bottom. Ma et al. \cite{mmoe} introduce the concept of sharing a universal expert across tasks, employing a gating network to ascertain the relevance of this universal expert to diverse tasks. Building on this, Tang et al. \cite{ple} propose the PLE, which explicitly delineates shared experts from task-specific ones, thereby mitigating detrimental interference between task-specific and task-shared insights.
Concerning multi-domain learning, Sheng et al. \cite{star} propose a STAR topology approach to model the shared centered parameters and domain-specific parameters among domains to keep each domain separate.
Nevertheless, in real-world business domains, frequent additions and updates to web pages are commonplace. The above-mentioned multi-domain models still rely on sufficient training data from the target domain and fail to alleviate the cold start problem of new domains.

If multi-domain models are employed, mixing data and networks from different domains can inadvertently lead to negative transfer of user preferences from the source domain to target domains, neglecting the specific recommendations required in the target domain. Consequently, these models exhibit limited efficacy in cold-start recommendations in new domains and fail to fully utilize the combination of target domain scenario information.

\section{Method}

\subsection{Problem Formulation}
\label{ssec:dynamic graph construction}
Typically, domains are divided into the source domain \(S\) and the target domain \(T\) according to general cross-domain recommendation settings. Each domain comprises a set of users \(U = \{u_1, u_2, ..., u_i\}\), a collection of items \(V = \{v_1, v_2, ..., v_j\}\), and labels \(y\). \(y \in \{0, 1\}\) indicates whether user \(u\) clicked on item \(v\) or not. 
For a user \(u_i\), he usually exhibit \(N\) distinct types of behaviors (e.g., browsing, clicking, purchasing) within a domain, which can be organized into a sequence of user behaviors \(B = \{b_1, b_2, \dots, b_N\}\). The multi-behavioral historical sequence for a user \(u\) is denoted by \(S_{u_i} = \{ s_{u_i}^{b_1}, s_{u_i}^{b_2}, \dots, s_{u_i}^{b_N}\}\), where \(s_{u}^{b}\) represents the sequence of behaviors under a specific action \(b\). For each behavior, there exists a sequential series of items \(s_{u}^{b} = \{v_{1}, v_{2}, \dots, v_{n}\}\), with \(n\) indicating the number of items interacted with.

To differentiate between the source and target domains, the user and item sets of the source domain are denoted as \(U^s\) and \(V^s\), respectively. In the case of the target domain \(D^t\), these are represented as \(U^t\) and \(V^t\). The embeddings for users \(u_i^d \in \mathbb{R}^k\) and items \(v_j^d \in \mathbb{R}^k\) are symbolized respectively as \(e_{u_i}^{d}\) and \(e_{v_j}^{d}\), where \(k\) indicates the dimensionality of the embeddings, and \(d \in \{s, t\}\) signifies the domain label. Additionally, \(e_d^{t}\) represents the embedding for the target domain special feature. 
In our problem setting, cold-start users are defined as those belonging to \(U^s\) but new to \(U^t\). The goal of CDR is to utilize the rich behavioral data from the source domain to provide recommendations for users \(u\) within the \(U^t\) set.

\subsection{Domain-Aware Cross-Attention}
\label{Domain-Aware Attention}
The framework of our method is shown in Fig \ref{fig:method}.
Transfer learning typically relies on the abundant explicit and implicit data from the source domain to train the transfer function, with the aim of mapping this embedding to the target domain. 
However, as previously mentioned, failing to fully consider the characteristics of the target domain may lead to noise and negative transfer problems when directly transferring user embeddings from the source domain to the target domain. 
Given the different characteristics and business scopes of different domains, users' behavioral preferences often diverge across these domains. 
Hence, it is crucial to consider the contribution of items from the user's source domain behavior during the cross-domain process.
Only an accurate evaluation of the contribution of the source domain can have a positive impact on the target domain recommendation results. 

Inspired by the Cross-Attention (CA) mechanism \cite{crossvit}, we construct an two-step domain-aware cross-attention network to fully consider such contributions from the source domain, thereby generating features that better match the target domain. 
We believe that both domain-level and item-level contributions should be considered when evaluating this part.
Cross-attention combines asymmetrically two separate embedding sequences of same dimension, in contrast self-attention input is a single embedding sequence. 
\begin{gather}
    \text{CA}(Q,K,V) = softmax(\frac{QK^T}{\sqrt{d_k}}) V, \\
    Q_i = XW_i^Q,
    K_i = X W_i^K,
    V_i=X W_i^V
\end{gather}
In this network, the first step involves a coarse-grained allocation of attention to the source domain's behavior sequence using the features of the target domain, and the second step refines the attention allocation by using the target domain item embeddings with the output of the first step. The two-step domain-aware cross-attention network can be expressed by the following formula: 
\begin{gather}
    e_1 = \text{Domain-Level CA}(XW_1^Q, e_d^tW_1^K, e_d^tW_1^V)\\
    e_z^t = \text{Item-Level CA}(e_1W_2^Q, e_{v_i}^tW_2^K, e_{v_i}^tW_2^V)
\end{gather}
where $X$ represents the behavior sequence feature $\{v_1, v_2, v_3, \dots \}$ of user $u_i$, 
$e_d^t$ represents target domain feature embedding, 
$e_{v_i}^t$ represents the target item embedding of $v_i$, 
$e_z^t \in \mathbb{R}^k$ denotes the transmissible feature embedding of user $u_i$ from source domain and 
$z$ represents the user behavior sequences and other side info sequence. 
In this way we can get transferable knowledge $e_{\text{seq}}^t$ and $e_{\text{side}}^t$ which contain rich source knowledge and are suitable for target domain.   
$W^Q \in \mathbb{R}^{d_1\times d_k}$, $W^K, W^V \in \mathbb{R}^{d_2\times d_k}$. 
It should be noted that this two-step domain-aware attention network assist in identifying source domain user interactions that are relevant to the target domain from coarse and fine granularity. 

\subsection{Domain Encoder}
\label{ssec:DADB}




In the cross-domain transfer process, the existing bridge-based methods \cite{EMCDR, SSCDR, ptuprefs4} train a designed Meta Network $f(\cdot)$, taking the outputs of the Cross Attention Network as Meta bridge parameters and the user embeddings from the source domain as input to calculate the transferred user embeddings:
\begin{equation}
    \hat{e}_{u_i}^t = f(e_{u_i}^s; w) 
\end{equation}
where $e_{u_i}^s$ denotes the embedding of user $u_i^s$ in source domain, $\hat{e}_{u_i}^t$ represents the transformed embedding, $w$ represents the parameters of the meta network. 

To realize personalized transfer meta network, some existing methods take the user sequence embedding as $w$. 
However, in pratical applications, user-side features typically encompass not only ID embeddings but also a large numbers of other auxiliary features. 
All these features will be fed into the final FC layers to predict the output of the target domain. 
When training with these auxiliary features, one problem is caused that the update gradients of the user embeddings obtained from the meta network are far smaller than those of the other auxiliary features. 
To address this issue, we adopt a domain encoder to generate transformed target user embeddings for simplicity instead of multiplying the behavior sequence feature with the user embedding. 
We flatten the output of two-step Domain-Aware Cross-Attention and concat them with each other to generate the transferred user embeddings. The proposed domain encoder network can be formalized as follows: 
\begin{equation}
    \hat{e}_{u_i}^t = g( concat[ e_{\text{seq}}^t, e_{\text{side}}^t, e_d^t ] )
\end{equation}
where $g(\cdot)$ represent the domain encoder, $e_d^t$ represent the target domain feature embedding and $e_{\text{side}}^t$ and $e_{\text{seq}}^t$ are the output embeddings from domain-aware cross-attention. 

Consequently, we directly obtain the transferred user embedding, avoiding the problem of  that the Meta Network may encounter during training. 


\subsection{Prediction and Optimization}
\label{pred}

With the generated target user embeddings, the output of the model can be formed as follows:
\begin{equation}
    \hat{y} = h(concat[\hat{e}_{u_i}^t, e_{\text{seq}}^t, e_{\text{side}}^t, e_{v_j}^t, e_d^t]; w)
\end{equation}
where $h(\cdot)$ represents the FC layers and $w$ represents the model parameters. 
In addition, we introduce a task-oriented optimization strategy inspired by work \cite{PTUPCDR}. This strategy does not directly minimize the distance between the target user embedding$u^t$ and the generated user embedding$\hat{u}^t$. Instead, but optimize the generated user embedding performance of the target domain. 
The optimization can be formulated as: 
\begin{equation}
    \mathcal{L} = \frac{1}{T} \sum_{i=1}^T - y_i log(\hat{y}_i - (1-y_i)log(1-\hat{y_i})
\end{equation}
where $y_i \in \{ 0, 1\}$ is the ground truth of the i-th sample, $\hat{y}_i$ is the predicted CTR value and T is the total sample number. 
This task-oriented optimization has two main advantages. It can mitigate the impact of insufficient target user embeddings, as it directly employs actual behavior label rather than approximate intermediate results. Through this optimization, the task-oriented learning process can utilize more training samples than comparing the distance between each other, thereby reducing the risk of overfitting. 

\subsection{Fine-tuning}
In our practical business situations, the addition of new domain is quite common.  Each domain has a dedicated item set and we need to make specific recommendations that are suitable for this domain within the dedicated item set. 
However, creating separate models for each new page would lead to severe cold-start issues and the complexity of engineering maintenance. 
Therefore, employing less model to serve different domains has become an urgent issue to address.

Our model DACDR is very friendly to the fine-tuning of new scenes. 
For the creation of new domains, we can fine-tune DACDR by training the target domain features $e_d^t$, target item embedding $e_{v_i}^t$ and  $W^Q, W^K, W^V$ in the two-step cross-attention part, and freeze the remaining network parameters. 
This allows for the rapid online deployment of the fine-tuned model.

\section{Experiments}
\label{Experiments}
In this section, we perform experiments on real-world datasets to evaluate the performance of our model. 

\subsection{Datasets and Implementation}
Experiments were conducted on both real-world industrial datasets and public datasets. 
The selection of the public datasets adhered to the choices made by the majority of the current methodologies \cite{SSCDR, TMCDR}, namely the Books, Movies and TV, CDs and Vinyl from Amazon\footnote{http://jmcauley.ucsd.edu/data/amazon/}. 
The industrial dataset is collected from our e-commerce online advertising system, encompassing a wealth of information across diverse domains. We extracted a subset of logs from September 1, 2023, to September 11, 2023, spanning 11 days, with the initial ten days allocated for training and the data from the final day reserved for testing purposes.

\begin{table}[ht]
\centering
\caption{Statistics of datasets used in experiments.}
\scalebox{0.85}{
\begin{tabular}{c|c|c|c|c} 
        \hline
        Datasets                & \#Users & \#Items   & \#Interactions    & CTR    \\
        \hline
        Source  &  19.9M & 32.8M   &  36B & 4.02\% \\
        Target  \#1 &  2.2M(94\% overlap)   & 210K    &   22.1M  & 1.48\% \\
        Target  \#2 &  4.5M(69\% overlap)    & 22K      &   39.1M   & 2.19\% \\
        Target  \#3 &  1.5M(78\% overlap)    & 491K      &   21.9M   & 1.85\% \\
        \hline
        \hline
        Movies & 123,960 & 50,052 & 1,697,533 & / \\
        CDs &  75,258 & 64,443 & 1,097,592 & / \\
        Books & 603,668 & 367,982 & 8,898,041 & / \\
        \hline
    \end{tabular}
    }
    \label{tab:statistics}
\end{table}

Within our business domains, the primary focus was directed towards data from two areas: central and target domains. Throughout the experiments, these were denoted as the source domain $S$ and the target domain $T$, respectively. The target domain typically has divergent business goals from the source domain. For instance, one user prefer to purchase more home appliances categories in the source domain, but shows more interest in clothing and shoes in domain \#1. 

Table \ref{tab:statistics} summarizes the statistical information of the public Amazon and our industrial datasets, detailing fundamental characteristics, the sparsity of each task within every domain, and the degree of overlap between users and exposed items across domains. Even though the domains share the same pool of items and contain numerous overlapping users, variations in item exposure and user behavior between the different domains were discernible. This suggests that users exhibit diverse behavioral intentions across multiple domains, engaging in a differentiated consumption ecosystem.

Our metric for evaluation was the AUC (Area under the ROC curve), the most commonly employed indicator in CTR prediction.



\subsection{Baselines and Metrics}
We compare our model with bellowing methods. 

\noindent$\bullet$ DNN \cite{youtubeDNN}: This method is a implement of YouTube DNN. In DNN-single, we train models for each domain separately using only target domain dataset, while in DNN-multi, we train models with datasets from both source domain and target domain.

\noindent$\bullet$ CMF \cite{CMF}: CMF is an extension of Matrix Factorization (MF), in which user embeddings are shared across the source and target domains.

\noindent$\bullet$ EMCDR \cite{EMCDR}: A  popular CDR method which employs MF to learn embeddings, and utilization of a neural network to bridge user embeddings from the auxiliary domain to the target domain.

\noindent$\bullet$ SSCDR \cite{SSCDR}: A semi-supervised bridge-based method for cross domain recommendation.

\noindent$\bullet$ Shared-Bottom \cite{ShareBottom}: Shared-Bottom is a classical multi-task method that consists of shared-bottom networks and several domain-specific tower networks. Each domain has its specific tower network while sharing the same bottom network.

\noindent$\bullet$ MMoE \cite{mmoe}: MMoE adopts the Mixture-of-Experts (MoE) structure by sharing the expert modules across all domains, while having a gating network trained for each domain.

\noindent$\bullet$ PLE \cite{ple}: PLE is optimized version of MMoE, which separates shared experts and task-specific experts explicitly and adopts a progressive mechanism to extract features gradually.

\noindent$\bullet$ STAR \cite{star}: is the state-of-the-art MDR method. It splits the parameters into shared and specific parts. Meanwhile, it proposes a Partitioned Normalization for distinct domain statistics.

\begin{table}
\centering
\caption{Overall Performance comparisons on our online production dataset w.r.t. AUC}
\scalebox{0.9}{
    \begin{tabular}{c|c|c|c} 
        \hline
        \multirow{2}*{Methods} & Target \#1 &  Target \#2  &  Target \#3 \\
                                                & AUC &  AUC   &   AUC   \\
        \hline
        DNN-single      & 0.6239 & 0.6746  & 0.6317\\
        CMF                     & 0.5828 & 0.6529  & 0.5965\\
        SSCDR               & 0.6136 & 0.6673  & 0.6085\\
        EMCDR               & 0.6076 & 0.6758  & 0.6244\\
        PTUPCDR         & 0.6329 & 0.6846  & 0.6363\\
        \hline
        DNN-multi       & 0.6336  & 0.6762 & 0.6479\\
        Share Bottom& 0.6401 & 0.6859  & 0.6563\\
        MMoE                & 0.6453  & 0.6878 & 0.6658\\
        PLE                     & 0.6489 & 0.6951  & 0.6597\\
        STAR                  & 0.6437 & 0.6956  & 0.6631\\
        \hline
        DACDR& \textbf{0.6513} & \textbf{0.7023}  & \textbf{0.6672} \\
        DACDR w/o DA       & 0.6498 & 0.6982 & 0.6631 \\
        DACDR w/o IA        & 0.6423 & 0.6938 & 0.6577 \\
        DACDR w/o DA\&IA  & 0.6401 & 0.6903 & 0.6518 \\
        \hline
    \end{tabular}
    }
    \label{tab:industrial exp}
\end{table}

Table \ref{tab:industrial exp} summarizes the performance of our proposed DACDR model and all the baselines on industrial datasets. The proposed DACDR model outperforms all the other baselines in all target domains. We observe that the majority of multi-task model results outperform those of cross-domain models. 
The reason for this is that multi-task models, through direct or indirect sharing of training embeddings or experts across different tasks, thus endow the models with a certain level of implicit understanding of the target domain. 
In contrast, our model, while considering the transfer of user preferences, has reinforced the model's capability to understand the target domain during the cross-domain process. 
Therefore, DACDR can exhibit superior performance.

DACDR has been deployed in our online recommender systems and achieves a 3.5\% increase in CTR and a 7.4\% increase in effective cost per mille (ECPM) in our business. It is remarkable that these advancements are substantial and may bring in millions of incomes within a month.
\begin{table}
\centering
\caption{Cold-start performance of 3 public cross-domain tasks w.r.t MAE and RMSE. }
\scalebox{0.63}{
    \begin{tabular}{c|c|c|c|c|c|c|c||c} 
        \hline
         & $\beta$ & Metric & CMF & SSCDR & EMCDR &  PTUPCDR & DACDR & Improve \\
        \hline
        \multirow{2}*{Movie} & \multirow{2}*{20\%} & MAE & 1.5209 & 1.3723 & 1.3507 & 1.1293 & \textbf{0.8265} & 26.81\% \\
                                                                                &  & RMSE & 2.0158 & 1.7304 & 1.6737 & 1.4734 & \textbf{1.1953} & 18.86\% \\ 
        \multirow{2}*{$\downarrow$} & \multirow{2}*{50\%} 
                                                                                        & MAE & 1.6893 & 1.5535 & 1.5312 & 1.2183 & \textbf{0.9327} & 23.45\% \\
                                                                                &  & RMSE & 2.2271 & 1.9032 & 1.8832 & 1.6388 & \textbf{1.4138} & 13.73\% \\
        \multirow{2}*{CDs} & \multirow{2}*{80\%} & MAE & 2.4186 & 2.0728 & 2.0441 &  1.5209 & \textbf{1.2420} & 18.36\% \\
                                                                                &  & RMSE & 3.0936 & 2.3490 & 2.4277 & 2.1238 & \textbf{1.8964} & 10.70\% \\

        \hline
        \multirow{2}*{Book} & \multirow{2}*{20\%} & MAE & 1.3632 & 1.1732 & 1.1328 & 1.0701 & \textbf{0.9042} & 15.50\% \\ 
                                                                                &  & RMSE & 1.7918 & 1.6327 & 1.4227 & 1.3698 & \textbf{1.1654} & 14.92\% \\
        \multirow{2}*{$\downarrow$} & \multirow{2}*{50\%} 
                                                                                        & MAE & 1.5813 & 1.3045 & 1.1836 & 1.0972 & \textbf{0.9123} & 16.86\% \\
                                                                                &  & RMSE & 2.0886 & 1.5737 & 1.4984 & 1.4356 & \textbf{1.2119} & 15.58\% \\
        \multirow{2}*{Movie} & \multirow{2}*{80\%} & MAE & 2.1577 & 1.3668 & 1.3545 &  1.2084 & \textbf{0.9879} & 18.24\% \\
                                                                                &  & RMSE & 2.6777 & 1.7198 & 1.7151  & 1.6135 & \textbf{1.3411} & 16.89\% \\
        \hline
        \multirow{2}*{Book} & \multirow{2}*{20\%} & MAE & 2.0084 & 1.9324 & 1.8951 & 1.4523 &  \textbf{0.9503} & 34.60\% \\
                                                                                &  & RMSE & 2.3829 & 2.3424 & 2.3163 & 1.9910 & \textbf{1.3800} & 30.66\% \\ 
        \multirow{2}*{$\downarrow$} & \multirow{2}*{50\%} 
                                                                                        & MAE & 2.1856 & 2.1417 & 2.1113 & 1.5445 & \textbf{0.9916} & 35.76\% \\
                                                                                &  & RMSE & 2.5373 & 2.4937 & 2.4681 & 2.0886 & \textbf{1.4391} & 31.08\% \\
        \multirow{2}*{CDs} & \multirow{2}*{80\%} & MAE & 2.8737 & 2.5126 & 2.4388 & 1.7968  & \textbf{1.1949} & 33.50\% \\
                                                                                &  & RMSE & 3.3424 & 2.9302 & 2.8110 & 2.4605 & \textbf{1.8245} & 25.82\% \\
        \hline
    \end{tabular}
    }
    \label{tab:public exp}
\end{table}

In our experiments, we test Cold-start performance on CDR public datasets. Following \cite{EMCDR, PTUPCDR} we set the proportions of cold-start users $\beta$ as 80\%, 50\%, and 20\% of the total overlapping users, respectively. 
Given that the ground truth is 5-core ratings, we modified the loss function to Mean Squared Error (MSE) Loss and updated the evaluation metrics to Mean Absolute Error (MAE) and Root Mean Squared Error (RMSE). The results are shown in Table \ref{tab:public exp}. 
Additionally, we take the sequential timestamps into account to avoid information leakage. Our method outperforms other bridge-based CDR methods on public datasets. We attribute this superiority to simplifying the three-step training process when training bridge-based methods with the meta network, opting to directly train a domain encoder instead of the meta method.

\subsection{Ablation Study}
To explore the effectiveness of key modules in our approach, we performed ablation experiments to verify the importance of our method DACDR by removing Domain-levels cross-Attention (w/o DA) and Item-level cross-Attention (w/o IA). The results are shown in Table \ref{tab:industrial exp}. 
Without domain-level attention, DACDR's performance drops but still outperforms most other methods. The same occurs with item-level attention. Removing both domain adaptation and instance adaptation simultaneously results in the most significant decrease in DACDR's performance, demonstrating the effectiveness of the attention module in CDR.

\section{Conclusion}
In this paper, we focus on the CDR problem and propose the DACDR model. We build a two-step domain-aware cross-attention network to capture coarse and fine-grained knowledge from the source domain. From DACDR, we observe the significant role of transformers in cross-domain recommendations, where domain information serves as a condition to guide the model towards better classification. In addition, our model can be easily fine-tuned and deployed, and is effective in handling new scenarios in practical applications. Experimental results on the industrial dataset and public dataset demonstrate the effectiveness of our method.


\bibliographystyle{IEEEbib}
\bibliography{refs}

\end{document}